\documentclass[11pt,a4paper]{article}
\usepackage{amsthm,fullpage}
\usepackage{xcolor}
\usepackage{graphicx}
\usepackage{multirow}
\newtheorem{theorem}{Theorem}
\newtheorem{lemma}[theorem]{Lemma}
\newtheorem{corollary}[theorem]{Corollary}
\newtheorem{definition}{Definition}

\newcommand{\SA}{\ensuremath{\mathrm{SA}}}
\newcommand{\LCP}{\ensuremath{\mathrm{LCP}}}
\newcommand{\BWT}{\ensuremath{\mathrm{BWT}}}
\newcommand{\LF}{\ensuremath{\mathrm{LF}}}

\newcommand{\tag}{\ensuremath{\mathrm{tag}}}
\newcommand{\Tag}{\ensuremath{\mathrm{Tag}}}
\newcommand{\XMS}{\ensuremath{\mathrm{XMS}}}
\newcommand{\len}{\ensuremath{\mathrm{len}}}
\newcommand{\ptr}{\ensuremath{\mathrm{ptr}}}
\newcommand{\pos}{\ensuremath{\mathrm{pos}}}
\newcommand{\rank}{\ensuremath{\mathrm{rank}}}
\newcommand{\TS}{\ensuremath{\mathrm{TS}}}
\newcommand{\run}{\ensuremath{\mathrm{run}}}
\newcommand{\up}{\ensuremath{\mathrm{up}}}
\newcommand{\down}{\ensuremath{\mathrm{down}}}
\newcommand{\triple}{\ensuremath{\mathrm{Triple}}}
\newcommand{\T}{\ensuremath{\mathcal{T}}}

\newcommand{\ag}[1]{{\color{blue!90!black}Adrian: #1}}

\newcommand{\no}[1]{}

\usepackage{authblk}

\author[1]{Andrej Bal\'az}
\author[2]{Travis Gagie}
\author[1]{Adri\'an Goga}
\author[3]{Simon Heumos}
\author[4]{Gonzalo Navarro}
\author[1]{Alessia	Petescia}
\author[5]{Jouni Sirén}
\affil[1]{\small Comenius University in Bratislava, Slovakia,
{\small \tt andrejbalaz001@gmail.com, adriangoga0@gmail.com, alessia.petescia@gmail.com} \medskip}
\affil[2]{CeBiB \& Dalhousie University, Canada,
{\small \tt travis.gagie@gmail.com}}
\affil[3]{University of Tuebingen, Germany,
{\small \tt simon.heumos@qbic.uni-tuebingen.de}}
\affil[4]{CeBiB \& DCC, University of Chile, Chile,
{\small \tt gnavarro@dcc.uchile.cl}}
\affil[5]{University of California Santa Cruz Genomics Institute, USA,
{\small \tt jltsiren@gmail.com}}

\begin{document}

\title{Wheeler maps}

\date{}

\maketitle

\begin{abstract}
Motivated by challenges in pangenomic read alignment, we propose a generalization of Wheeler graphs that we call Wheeler maps. A Wheeler map stores a text $T[1..n]$ and an assignment of tags to the characters of $T$ such that we can preprocess a pattern $P[1..m]$ and then, given $i$ and $j$, quickly return all the distinct tags labeling the first characters of the occurrences of $P[i..j]$ in $T$. For the applications that most interest us, characters with long common contexts are likely to have the same tag, so we consider the number $t$ of runs in the list of tags sorted by their characters' positions in the Burrows-Wheeler Transform (BWT) of $T$. We show how, given a straight-line program with $g$ rules for $T$, we can build an $O(g + r + t)$-space Wheeler map, where $r$ is the number of runs in the BWT of $T$, with which we can preprocess a pattern $P[1..m]$ in $O(m \log n)$ time and then return the $k$ distinct tags for $P[i..j]$ in optimal $O(k)$ time for any given $i$ and $j$. We show various further results related to prioritizing the most frequent tags. 
\end{abstract}

\newpage 

\section{Introduction}
\label{sec:introduction}

For years, geneticists have been worried about the fact that using a single reference for the human genomes biases scientific studies and medical diagnoses, undermining the potential of personalized medicine, particularly for people from under-represented groups. To address this bias, researchers~\cite{liao2023hprc} recently published a pangenome consisting of nearly complete genomes from 47 people from diverse origins and took, according to the {\it New York Times}, ``a major step toward a deeper understanding of human biology and personalized medicine for people from a wide range of racial and ethnic backgrounds''. Eventually, the plan is to include 350 genomes, but even this many genomes cannot fully capture humanity's genetic diversity. As the {\it Guardian} put it, ``as long as the reference contains only a subset, arguably someone will not make the cut''. Ultimately, there will be pressure for a reference of at least thousands of genomes.

One of the primary use of a reference is during the read alignment. As a DNA sample passes through a sequencing machine, the machine records the genome in short substrings called {\em reads}. The length and accuracy of the reads vary depending on the sequencing technology used. Next, software called read aligner uses an index of a reference to find {\em seeds}, sections of the reads that exactly match sections in the reference, and uses dynamic programming to extend those seeds to approximate matches of the whole read. These approximate matches form alignments, which are used in many subsequent bioinformatics analyses.

Indexing 47 human genomes is feasible even with standard read aligners such as Bowtie~\cite{langmead2012fast} and BWA~\cite{li2009fast}, and even indexing 350 may be possible on supercomputers, but indexing thousands will require new data structures. The emerging consensus is that we should represent the combined reference sequences as a {\em pangenome graph} that shows variation between genomes as detours on an otherwise shared path. The necessity of mapping reads to the version of the path that best fits the sample leads to the question of how to index pangenome graphs.

Equi et al.~\cite{equi2021indexing} showed that, unless the strong exponential-time hypothesis is false, one cannot index a graph in polynomial time such that pattern matching can run in sub-quadratic time, so several groups have tried constraining pangenome graphs to have a particular structure, such as Wheeler graphs \cite{gagie2017wheeler}, $p$-sortable graphs \cite{cotumaccio2021indexing}, elastic degenerate strings \cite{bernardini2017pattern} or founder block graphs \cite{makinen2020founder}`. Unfortunately, these constraints may not reflect biological models realistically, and even if they do, merging reference sequences into a graph hides certain variations' tendency to co-occur, known as {\em linkage disequilibrium}~\cite{reich2001linkage}, and creates {\em chimeric} paths whose labels are not in any of the original sequences.  Indexing and using such a graph can result in false-positive matches to these chimeric paths.

Some groups reduce the number of false positives by excluding rare variations since the more variations are represented, the noisier the graph becomes and the more possibilities there are for spurious matches. Sacrificing inclusivity for the sake of computational convenience 
goes against the spirit of pangenomics, and the pressure to include more genomes will probably force bioinformaticians to index all the variations. Moreover, excluding variations could be viewed as trading false positives for false negatives. Other groups filter out false positives by checking matches against the reference sequences represented as strings. Still, their overall query time cannot be bounded in terms of the patterns and the true matches reported. Furthermore, the number of false positives will likely grow as the pangenome does.

Yet other researchers have eschewed using a pangenome graph altogether and indexed the genomes in the pangenome as a set of strings. This approach allowed them to draw on a rich history of indexing highly repetitive texts: the Burrows-Wheeler Transform (BWT) and FM-indexes, for which Burrows, Ferragina and Manzini recently shared the Paris Kanellakis Award and which underpin Bowtie and BWA; RLCSA; the r-index \cite{gagie2018optimal}, subsampled r-index \cite{cobas2021subsampled} and r-index-f \cite{tatarnikov_et_al2023monik}. 
Recently, Rossi et al.~\cite{rossi2022moni} and Boucher et al.~\cite{boucher2021phoni} showed how, given a straight-line program with $g$ rules for a text $T[1..n]$, they can build an $O(g + r)$ space index, where $r$ is the number of runs in the BWT of $T$, with which they can find the maximal exact matches (MEMs) of a given pattern $P[1..m]$ with respect to $T$ in $O(m \log n)$ time and list the occurrences of each MEM in constant time per occurrence. This result means they can index the pangenome compactly with no chance of false positives, find good seeds reasonably quickly, and list the occurrences of those seeds in constant time per occurrence.

The main practical problem with Rossi et al.'s result is that if there are thousands of genomes in the pangenome, then a MEM could occur thousands of times in those genomes, even if all those occurrences map to only one place in the standard single reference genome. This observation makes extending the seeds and combining the approximate matches of the reads much more difficult. In this paper, we show how we can combine Rossi et al.'s result with a pangenome graph such that we can still find seeds quickly, with no chance of false positives, but then report their non-chimeric occurrences in the graph in constant time per occurrence. Moreover, we put no constraints on the graph.

A set of genomes can be annotated so that for each character in the genomes, we know at which vertex in a pangenome graph that character occurs.
Then, our idea is that if someone gives us a set of genomes and the corresponding annotation, we can store them in a small space so we can later quickly report for each seed its starting positions in the pangenomic graph.
The seeds of a read with respect to the set of genomes can be MEMs, but also $f$-MEMs~\cite{navarro2016compact} (maximal substrings that occur at least $f$ times in the genomes) or other kinds of substrings. 

We can formalize this problem as follows: we want to store a text $T[1..n]$ and an assignment of tags to the characters of $T$ such that we can preprocess a pattern $P[1..m]$ and then, given $i$ and $j$, quickly return all the distinct tags labeling the first characters the occurrences of $P[i..j]$ in $T$. In a pangenome, characters with long common contexts are more likely to have the same tag, so we consider the number of runs $t$ in the list of tags sorted by their characters' positions in the Burrows-Wheeler Transform (BWT) of $T$.

\paragraph*{Our contribution.}
In this paper, we show how, given an SLP with $g$ rules for $T$, we can build an $O(g + r + t)$ space data structure, where $r$ is the number of runs in the BWT of $T$, with which we can preprocess a pattern $P$ in $O(m \log n)$ time and then return the $k$ distinct tags for $P[i..j]$ in optimal $O(k)$ time for any given $i$ and $j$. Without raising the space occupied by our data structure or the time it takes to preprocess $P$, we present further results on prioritizing and constraining the query tag frequencies.  
In more detail, we show how to count the number of distinct tags labeling the occurrences of $P[i..j]$ in constant time and return the top $k$ most frequent tags among them in $O ((\log t + k) \log^\epsilon t)$, for any $\epsilon > 0$. As our final contribution, we propose an extension of the notion of $f$-MEMs \cite{navarro2016compact} to tags: given a parameter $f$ fixed at construction time, after preprocessing $P$, we can later report the $k$ resulting tags that label at least $f$ occurrences of $P[i..j]$ in $O ((k+1)\log^\epsilon t)$ time for any $\epsilon > 0$.

We call our data structure a {\em Wheeler map} since it is something like a Wheeler graph~\cite{gagie2017wheeler} but with less structure.  One reason Wheeler graphs were introduced was to provide a model for alignment with a pangenome graph: we start with a string dataset, build a graphical representation, and index that graph; the graphical representation is inherently lossy but, to filter out chimeric matches, we can verify matches against the original dataset.  (Even before Wheeler graphs were defined, software for indexing variation graphs~\cite{vg} used a procedure for making them Wheeler or almost Wheeler, falling back on unwinding the graph and indexing substrings when that procedure failed.)  Our idea is to reverse that approach of indexing a graph and then filtering out false positives using the strings.  Instead, we index the strings, and then map occurrences onto a graph --- but without considering all the occurrences in the strings.

Some researchers argue that having a graph index return matches not found in the original strings is a feature, not a bug, since it allows the index to find matches that can be obtained by recombination.  We believe, however, that as computer scientists it is not our job to decide what combination of alleles are reasonable and we should index the dataset we are given and nothing else.  Indexing the strings means we index all the variations they contain, so we can presumably capture most reasonable combinations by increasing the number of genomes in our dataset.  Scaling to larger datasets is thus a solution for us, whereas it is a problem for graphical indexes, which tend to produce more false positives when they include all the variations in large datasets.

From Rossi et al.~\cite{rossi2022moni}, we know $r$ and $g$ are reasonably small for the datasets in which we are most interested.  To check that $t$ is comparable, we computed it for the chromosome-19 component in a Minigraph-Cactus graph based on 90 human haplotypes from the Human Pangenome Reference Consortium~\cite{liao2023hprc}.  This component was built from 1100 contigs with total length $n = 5{,}070{,}072{,}154$ and $t$ was $208{,}649{,}680$, almost 25 times smaller than $n$.  For comparison, $r$ was $71{,}512{,}609$, just over 70 times smaller than $n$ and not quite 3 times smaller than $t$.

\paragraph*{Roadmap.}
In Section \ref{sec:preliminaries} we describe the basic concepts that will be used throughout the rest of the work, together with a preliminary method of computing the tags for the occurrences of a pattern $P$. In Section \ref{sec:computing} we show how extended matching statistics can be computed $O (m \log n)$ time without the need for buffering that Rossi et al.\  \cite{rossi2022moni} used, and extend the method for computing the tag statistics. In Section \ref{sec:using} we describe how the tag statistics together with range successor queries on the tag array can be used to get the $k$ distinct tags for the occurrences of $P[i..j]$ in $O (\log^\epsilon t + k)$ time for any $\epsilon > 0$. Using more sophisticated techniques, we improve this time to the optimal $O (k)$ in Section \ref{sec:optimal}. We continue by showing additional results about constraining and prioritizing occurrence frequencies. Section \ref{sec:topk} shows how to count how many distinct tags label occurrences of $P[i..j]$ and how to report $k$ tags that label the most occurrences of $P[i..j]$. Section \ref{sec:fmems} describes how to efficiently obtain the $k$ tags that label at least $f$ occurrences of $P[i..j]$ in $T$. We conclude in Section~\ref{sec:concl} with some future work directions.

\section{Preliminaries}
\label{sec:preliminaries}
Our model of computation throughout is the standard word-RAM with $\Theta (\log n)$-bit words.

For the sake of brevity, we assume the reader is familiar with suffix arrays (SAs), the Burrows-Wheeler Transform (BWT), FM-indexes, LF-mapping and straight-line programs (SLPs); otherwise, we refer them to appropriate surveys~\cite{NMacmcs06,Navacmcs20.2}.  We remind only that $\LCP (S_1, S_2)$ denotes the length of the longest common prefix of two strings $S_1$ and $S_2$ (which need not be lexicographically consecutive suffixes of a text), and of the bounds for Muthukrishnan's~\cite{muthukrishnan2002efficient} classic document-listing data structure:

\begin{theorem}[Muthukrishnan, {\cite[Thm. 3.1]{muthukrishnan2002efficient}}]
\label{thm:Muthu}
Given an array $A [1..n]$, we can build an $O (n)$-space data structure with which, given $i$ and $j$, we can return the $k$ distinct elements in $A [i..j]$ in $O (k)$ time.
\end{theorem}


\begin{figure}[t]
\begin{center}
{\begin{tabular}{c@{\hspace{2ex}}c@{\hspace{2ex}}c}
rows 1--15 & rows 16--30 & rows 31--45 \\
\\
\begin{tabular}{c@{$~\!$}crrl}
\multicolumn{2}{c}{$L$} & tag & bwt & context\\
\hline
$^{^0}$ & & 9 & \tt T & \tt $\mathtt{\$}$AGATACA\\
     & & 9 &\tt T& \tt $\mathtt{\$}$GATACA\\
& $^{^0}$ & 9 & \tt T & \tt $\mathtt{\$}$GATTACA\\
$_{_0}$ & & 9 & \tt T & \tt $\mathtt{\$}$GATTAGA\\
    & $^{_0}$ & 10 &\tt A& \tt $\mathtt{\$}$GATTAGAT\\
$^{^0}$ & $^{_1}$ & 9 &\tt T& \tt A$\mathtt{\$}$GATTAGA\\
$^{^1}$ & & 4 & \tt T & \tt ACAT$\mathtt{\$}$AGA\\
  & $^{^4}$ & 4 &\tt T& \tt ACAT$\mathtt{\$}$GA\\
$^{^4}$ && 5 & \tt T & \tt ACAT$\mathtt{\$}$GAT\\
  & $^{_1}$& 5 &\tt T& \tt AGAT$\mathtt{\$}$GAT\\
&& 5 &\tt T& \tt AGATA$\mathtt{\$}$GAT\\
$^{^5}$ & $^{_8}$ & 0 & \$ & \tt AGATACAT\\
$^{^1}$ && 7 & \tt C & \tt AT$\mathtt{\$}$AGATA\\
&& 7 &\tt C& \tt AT$\mathtt{\$}$GATA\\
& $^{_2}$& 7 & \tt C & \tt AT$\mathtt{\$}$GATTA
\end{tabular}
&
\begin{tabular}{c@{$~\!$}crrl}
\multicolumn{2}{c}{$L$} & tag & bwt & context\\
\hline
&& 7 &\tt G& \tt AT$\mathtt{\$}$GATTA\\
&& 7 &\tt G& \tt ATA$\mathtt{\$}$GATTA\\
$^{^3}$&& 2 &\tt G& \tt ATACAT$\mathtt{\$}$A\\
&& 2 & \tt G& \tt ATACAT$\mathtt{\$}$\\
& $^{_2}$& 2 & \tt G & \tt ATTACAT$\mathtt{\$}$\\
&& 2 &\tt G& \tt ATTAGAT$\mathtt{\$}$\\
&& 2 &\tt G& \tt ATTAGATA$\mathtt{\$}$\\
$^{^0}$ && 6 & \tt A & \tt CAT$\mathtt{\$}$AGAT\\
&& 6 &\tt A& \tt CAT$\mathtt{\$}$GAT\\
& $^{_0}$ & 6 &\tt A & \tt CAT$\mathtt{\$}$GATT\\
&& 6 &\tt A& \tt GAT$\mathtt{\$}$GATT\\
&& 6 &\tt A& \tt GATA$\mathtt{\$}$GATT\\
$^{^4}$ && 1 &\tt A& \tt GATACAT$\mathtt{\$}$\\
&& 1 & \$ & \tt GATACAT\\
$_{_{\,}}$ & $^{_3}$ & 1 & \$ & \tt GATTACAT
\end{tabular}
&
\begin{tabular}{c@{$~\!$}crrl}
\multicolumn{2}{c}{$L$} & tag & bwt & context\\
\hline
&& 1 & \$ & \tt GATTAGAT\\
&& 1 & \$ & \tt GATTAGATA\\
$^{^0}$&& 8 & \tt A & \tt T$\mathtt{\$}$AGATAC\\
&& 8 &\tt A& \tt T$\mathtt{\$}$GATAC\\
& $^{_1}$& 8 & \tt A & \tt T$\mathtt{\$}$GATTAC\\
&& 8 & \tt A & \tt T$\mathtt{\$}$GATTAG\\
&& 8 &\tt A& \tt TA$\mathtt{\$}$GATTAG\\
$^{^2}$&& 3 & \tt A & \tt TACAT$\mathtt{\$}$AG\\
&$^{^5}$& 3 & \tt A & \tt TACAT$\mathtt{\$}$G\\
$^{^5}$&& 4 & \tt T & \tt TACAT$\mathtt{\$}$GA\\
&$^{_2}$ & 4 &\tt T& \tt TAGAT$\mathtt{\$}$GA\\
&& 4 &\tt T& \tt TAGATA$\mathtt{\$}$GA\\
$^{^1}$&& 3 & \tt A & \tt TTACAT$\mathtt{\$}$G\\
& $^{_3}$& 3 &\tt A& \tt TTAGAT$\mathtt{\$}$G\\
$_{_0}$&& 3 &\tt A& \tt TTAGATA$\mathtt{\$}$G
\end{tabular}
\end{tabular}}

\vspace{5ex}

\includegraphics[width=0.6\textwidth]{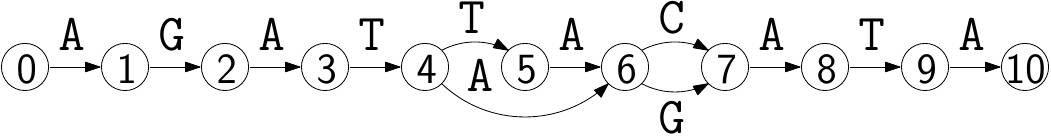}
\end{center}


\caption{On the bottom, an alignment graph for a set of (toy) genomes, concatenated in the text $T = {\tt GATTACAT\mathtt{\$}AGATACAT\mathtt{\$}GATACAT\mathtt{\$}GATTAGAT\mathtt{\$}GATTAGATA\mathtt{\$}}$. The tags are the identifer of the node the edge labeled by each symbol departs from. On the top, some arrays split in three parts. The tag array is shown in the middle columns. The right columns show the text contexts $T[\SA[i..]]$. The left columns show the entries of $L$ (leftward, inter-run LCPs, and rightward, intra-run LCPs). LCPs extend up to the terminators \$ because they are not searchable.}
\label{fig:rangesuccessor}
\end{figure}

In our model, each text suffix $T[i..]$ is labeled with a ``tag'', which can also be seen as labeling the position $i$. The tags of $T$ are collected in a so-called ``tag array'' (see Figure~\ref{fig:rangesuccessor}).

\begin{definition}
The tag array $\Tag[1..n]$ of a text $T[1..n]$ whose suffixes $T[i..]$ are labeled satisfies that $\Tag [q]$ is the label of suffix $T [\SA [q]..n]$.
\end{definition}

We say that an occurrence $T[i..i+|P|-1]$ of a pattern $P$ in $T$ is labeled by the tag that labels $T[i..]$. Consequently, the labels of all the occurrences of $P$ in $T$ are listed in $\Tag [s..e]$, where $\SA [s..e]$ is the suffix array interval for $P$. For example, in Figure~\ref{fig:rangesuccessor} the range for $P=\texttt{A}$ is $\SA[6..22]$, and $\Tag[6..22]$ contains the tags 9, 4, 5, 0, 7, and 2. Those are the labels with which $P$ appears in the graph.

For convenience, we first extend the standard definition of matching statistics to include the lexicographic ranks of the suffixes of $T$ starting with the occurrences we consider, and then further extend it to mention the tag array.

\begin{definition}
\label{def:XMS}
The {\em extended matching statistics} of a pattern $P [1..m]$ with respect to a text $T [1..n]$ are an array $\XMS [1..m+1]$ of $(\len, \pos, \rank)$ triples such that
\begin{itemize}
\item $\XMS [i].\len$ is the length of the longest prefix of $P [i..m]$ that occurs in $T$,
\item $\XMS [i].\pos$ is the starting position of one occurrence of $P [i..i + \XMS [i].\len - 1]$ in $T$,
\item $\XMS [i].\rank$ is the lexicographic rank of $T [\XMS [i].\pos..n]$ among the suffixes of $T$.
\end{itemize}
\end{definition}

We emphasize that we expect the tag array to have (hopefully few) runs of equal consecutive symbols. The following definition considers those runs in the process of matching $P$ in $T$.

\begin{definition}
\label{def:TS}
The {\em tag statistics} of a pattern $P [1..m]$ with respect to a text $T [1..n]$ and its tag array $\Tag [1..n]$  are an array $\TS [1..m+1]$ of $(\len, \pos, \rank, \run, \up, \down)$ sextuples such that $\TS [i].\len$, $\TS [i].\pos$ and $\TS [i].\rank$ are the same as in the $\XMS$ array and
\begin{itemize}
\item $\TS [i].\run$ is the index of the run $\Tag[u..d]$ in the tag array that contains position $\TS [i].\rank$,
\item $\TS [i].\up = \LCP (P [i..m], T [\SA [u]..n])$,
\item $\TS [i].\down = \LCP (P [i..m], T [\SA [d]..n])$.
\end{itemize}
\end{definition}

We rely on results about straight-line programs (SLPs), which we can encapsulate in the following lemma.

\begin{lemma}
\label{lem:LCP}
Given an SLP with $g$ rules for $T [1..n]$, in $O(n\log n)$ expected time 
we can build an $O (g)$-space data structure with which we can preprocess any pattern $P [1..m]$ in $O (m)$ time such that later, given $i$, $j$ and $q$, we can return $\LCP (P [i..j], T [q..n])$ in $O (\log n)$ time and with no chance of error as long as $P [i..j]$ occurs somewhere in $T$.
\end{lemma}

\begin{proof}
Bille et al.~\cite{bille2017fingerprints} showed how to build, in $O(n\log n)$ expected time, a Karp-Rabin hash function with no collisions between substrings of $T$. If $S = S' \cdot S''$ and we have the hashes of two of those strings, we can compute the hash of the third in constant time, as soon as we store some precomputed values that can also be maintained in constant time (see, e.g., \cite{NP18}).
 
If necessary, we use Ganardi et al.'s~\cite{ganardi2021balancing} construction to balance the SLP such that it has $O (g)$ rules and height $O (\log n)$. We then label each symbol $x$ in the SLP with the length and hash of $x$'s expansion. This takes $O(g)$ time because we compute in constant time the hash of the left-hand side of a rule from those of the right-hand side.

When $P$ arrives, we compute the hashes of its suffixes in $O(m)$ total time. The hash of any $P[i..j]$ can then be computed in constant time from the hashes of $P[i..m]$ and $P[j+1..m]$.

Given $i$, $j$ and $q$, we descend to the $q$th leaf of the parse tree in $O (\log n)$ time.  We then re-ascend toward the root in $O (\log n)$ time, keeping track of the length and hash of $T [q..e]$, where $e$ is the index of the rightmost leaf in the subtree of the node we are currently visiting.

When we reach a node such that $T [q..e]$ is either longer than $P [i..j]$ or the hash of $T [q..e]$ does not match the hash of the corresponding prefix of $P [i..j]$, we re-descend in $O (\log n)$ time.   At each step in the re-descent, we go left if $T [q..e]$ is either longer than $P [i..j]$ or the hash of $T [q..e]$ does not match the hash of the corresponding prefix of $P [i..j]$, where $e$ is now the index of the rightmost leaf in the subtree of the left child. Otherwise, we go right.

We then find $\LCP (P [i..j], T [q..n])$ in $O (\log n)$ time.  As long as $P [i..j]$ occurs somewhere in $T$, no hash of a prefix of $P [i..j]$ collides with the hash of a different substring of $T$, so we have no chance of error.
\end{proof}

We note as an aside that, since Navarro et al.~\cite{navarro2022balancing} recently extended Ganardi et al.'s balancing construction to run-length SLPs\footnote{Those allow rules of the form $A \rightarrow B^k$ for any $k > 0$, which count as constant-size. There are string families for which the smallest RLSLP is $\Theta(\log n)$ times smaller than the smallest SLP \cite{navarro2022balancing,Navacmcs20.3}.}, Lemma~\ref{lem:LCP} and all our results hold for those as well.

\label{sec:previous}

\medskip


Finally, although we know of no previous work specifically addressing tag arrays, it would be disingenuous of us not to state a result that follows from work by M\"akinen et al.~\cite{makinen2010storage}:

\begin{theorem}[M\"akinen et al., {\cite[Thm 17.]{makinen2010storage}}]
\label{thm:RLBWT}
Given a text $T [1..n]$ whose BWT has $r$ runs, we can build an $O (r)$-space data structure called RLBWT such that later, given a pattern $P [1..m]$, we can return the lexicographic range of suffixes of $T$ starting with $P$ in $O (m \log \log n)$ time.
\end{theorem}

\begin{corollary}
\label{cor:RLBWT}
Given a text $T [1..n]$ whose BWT has $r$ runs, and a tag array with $t$ runs, we can build an $O (r + t)$-space data structure such that later, given a pattern $P [1..m]$, we can return the $k$ distinct tags of $P$'s occurrences in $T$ in $O (m \log \log n + k)$ time.
\end{corollary}

\begin{proof}
We store an $O (r)$-space RLBWT for $T$, an $O (t)$-space predecessor structure storing 
where the runs start in $\Tag$, and an $O (t)$-space instance of Muthukrishnan's data structure from Theorem~\ref{thm:Muthu} for the array $A [1..t]$ obtained from $\Tag$ by replacing each run by a single copy of the same tag.
Given $P$, we first use the RLBWT to find the lexicographic range $\SA [s..e]$ of suffixes of $T$ starting with $P$, in $O (m \log \log n)$ time.  We then use predecessor queries 
to find the range $A [s'..e']$ of the tag run indices overlapping $\Tag [s..e]$, in $O (\log \log n)$ time.  Finally, we use Muthukrishnan's data structure to report the distinct tags in $A [s'..e']$, in $O (k)$ time.
\end{proof}

Our main concern with Corollary~\ref{cor:RLBWT} is that if we want the distinct tags for a set of substrings of $P$ that can overlap---such as the maximal exact matches (MEMs) of $P$ with respect to $T$---and we apply this corollary to each one, then we can use $\Omega (m^2)$ total time even when the number of tags we return is small. Our plan is then to preprocess $P$ in a first stage, so that in a second stage we can more quickly answer (many) questions about substrings of the form $P[i..j]$.

\section{Computing tag statistics}
\label{sec:computing}

We now describe our preprocessing of the pattern.
Our results in this section can be viewed as mainly extending Rossi et al.'s~\cite{rossi2022moni} work on computing (extended) matching statistics to computing tag statistics:

\begin{theorem}[cf.~\cite{rossi2022moni}]
\label{thm:MONI}
Given an SLP with $g$ rules for a text $T [1..n]$ whose BWT has $r$ runs, we can build an $O (g + r)$-space data structure such that later, given a pattern $P [1..m]$, we can compute the extended matching statistics $\XMS$ 
of $P$ with respect to $T$ in $O (m \log n)$ time.
\end{theorem}

\begin{proof}
We apply Lemma~\ref{lem:LCP} to the SLP to obtain an $O (g)$-space LCP data structure with $O (\log n)$ query time. We also store $\SA [u]$ and $\SA [d]$, for each run $\BWT [u..d]$, in an $O(r)$-space data structure supporting predecessor and successor queries on the keys $u$ and $d$. Finally, we use the $O(r)$-space RLBWT of Theorem~\ref{thm:RLBWT}, which can also compute any $\BWT[j]$ and $\LF[j]$. These functions and the predecessor queries can run in $O(\log\log n)$ time, but $O(\log n)$ time is enough for our purposes. 

As usual, for technical convenience we add to $T$ a special symbol $T[n+1] = \$$ that is lexicographically smaller than all the other symbols in $T$ (and in potential patterns $P$). This implies $\BWT[1]=\$$. For a start, then, considering $P[m+1..m] = \epsilon$, we set $\XMS [m+1].\len = 0$, $\XMS [m+1].\rank = 1$ and $\XMS [m+1].\pos = n+1$.

Now, suppose  we have already computed the suffix $\XMS [i + 1..m+1]$ of the extended matching statistics and want to compute $\XMS [i]$.  If $\BWT [\XMS [i + 1].\rank] = P [i]$ then
\begin{eqnarray*}
\XMS [i].\len & = & \XMS [i + 1].\len + 1\,,\\
\XMS [i].\pos & = & \XMS [i + 1].\pos - 1\,,\\
\XMS [i].\rank & = & \LF [\XMS [i + 1].\rank]\,.
\end{eqnarray*}
Otherwise, let $\BWT [u]$ and $\BWT [d]$ 
be the occurrences of $P [i]$ immediately preceding and following $\BWT [\XMS [i + 1].\rank]$. We find $u$ and $d$ with predecessor/successor queries.

By the definition of the BWT, at least one of $T [\SA [u]..n]$ and $T [\SA [d]..n]$ has the longest common prefix with $P [i + 1..m]$ of any suffix of $T$ preceded by a copy of $P [i]$.  Since $\BWT [u]$ is the last character in a run and $\BWT [d]$ is the first character in a run, we have $\SA [u]$ and $\SA [d]$ stored.  Therefore, we can compute
\begin{eqnarray*}
\ell_u & = & \LCP (P [i + 1..i + \XMS [i + 1].\len - 1], T [\SA[u]..n])\,,\\
\ell_d & = & \LCP (P [i + 1..i + \XMS [i + 1].\len - 1], T [\SA [d]..n])\,,
\end{eqnarray*}
in $O (\log n)$ time, since $P [i + 1..i + \XMS [i + 1].\len - 1]$ occurs in $T$, with no chance of error.

If $\ell_u \geq \ell_d$ then
\begin{eqnarray*}
\XMS [i].\len & = & \ell_u + 1\,,\\
\XMS [i].\pos & = & \SA [u] - 1\,,\\
\XMS [i].\rank & = & \LF [u]\,,
\end{eqnarray*}
and, symmetrically, if $\ell_u < \ell_d$ then
\begin{eqnarray*}
\XMS [i].\len & = & \ell_d + 1\,,\\
\XMS [i].\pos & = & \SA [d] - 1\,,\\
\XMS [i].\rank & = & \LF [d]\,.
\end{eqnarray*}
\end{proof}

\begin{corollary}
\label{cor:MONI}
Suppose we are given an SLP with $g$ rules for a text $T [1..n]$ whose $\BWT$ has $r$ runs, and a tag array for $T$ with $t$ runs.  Then we can build an $O (g + r + t)$-space data structure such that later, given a pattern $P [1..m]$, we can compute the tag statistics of $P$ with respect to $T$ in $O (m \log n)$ time.
\end{corollary}

\begin{proof}
We store an $O(t)$-space predecessor data structure on the starting positions of the runs in $\Tag$. For each run $\Tag[u..d]$, we also store $\SA [u]$ and $\SA [d]$. Given $P$, we start by applying Theorem~\ref{thm:MONI} to compute the extended matching statistics $\XMS [1..m+1]$ of $P$ with respect to $T$ in $O (m \log n)$ time.  For $1 \leq i \leq m+1$, we then set
\begin{eqnarray*}
\TS [i].\len & = & \XMS [i].\len\,,\\
\TS [i].\pos & = & \XMS [i].\pos\,,\\
\TS [i].\rank & = & \XMS [i].\rank\,,
\end{eqnarray*}
and $\TS [i].\run$ to the index of the run $\Tag[u..d]$ in the tag array containing position $\TS [i].\rank$ (computed with a predecessor query). Further, we use the LCP data structure to compute
\begin{eqnarray*}
\TS [i].\up & = & \LCP (P [i..m], T [\SA [u]..n])\,,\\
\TS [i].\down & = & \LCP (P [i..m], T [\SA [d]..n])\,.
\end{eqnarray*}
This also takes a total of $O (m \log n)$ time.
\end{proof}

\section{Using tag statistics}
\label{sec:using}


Once we have the tag statistics of $P$ with respect to $T$, we no longer need Lemma~\ref{lem:LCP}, or even the SA samples or BWT, to find out which tags label the occurrences of any $P[i..j]$.  We use Muthukrishnan's document-listing data structure in the same way as in the proof of Corollary~\ref{cor:RLBWT}: once we know which runs in the tag array overlap the BWT interval for $P [i..j]$, we use Muthukrishnan's structure to list the $k$ distinct tags in $O (k)$ time.  In this section we explain how we find which runs in the tag array overlap the BWT interval for $P[i..j]$, without computing the interval itself (which we do not know how to do quickly in $O (g + r + t)$ space).

Let $U [1..t]$ and $D [1..t]$ be the arrays such that $U [q]$ and $D [q]$ are the indices of the first and last tags, respectively, in the $q$th run in the tag array.  Let $W [1..t]$ be the array with
\[W [q]
= \min_{U [q] + 1 \leq p \leq D [q]} \{ \LCP (T [\SA [p - 1]..n], T [\SA [p]..n]) \}
= \LCP(T[\SA[U[q]]..n],T[\SA[D[q]]..n])\]
for $1 \leq q \leq t$, and let $B [1..t - 1]$ be the array with
\[B [q] ~=~ \LCP (T [\SA [D [q]]..n], T [\SA [U [q + 1]]..n])\]
for $1 \leq q \leq t - 1$---so $W [q]$ is the LCP computed {\em within} run $q$ and $B [q]$ is the LCP computed {\em between} runs $q$ and $q + 1$. Finally, let
\[L [0..2 t] ~=~ 0, W [1], B [1], W [2], B [2], \ldots, W [t - 1], B [t - 1], W [t], 0\,.\]
We will use $L$ in this section and the next one, but we will not refer to $U$, $D$, $W$ or $B$ again. We remind that Figure~\ref{fig:rangesuccessor} shows the $L$ array on an example text.

\begin{lemma}
\label{lem:range-successor}
Suppose we are given a text $T [1..n]$ and a tag array for $T$ with $t$ runs. Then, for any constant $\epsilon>0$, we can build an $O (g+t)$-space data structure such that later, given the tag statistics of a pattern $P [1..m]$ with respect to $T$ and $i$ and $j$, we can find which runs in the tag array overlap the BWT interval for $P [i..j]$ in $O (\log^\epsilon t)$ time.
\end{lemma}

\begin{proof}
We store $O (t)$-space range-predecessor/successor data structures over $L$ with $O (\log^\epsilon t)$ query time~\cite{nekrich2012sorted} (we call them collectively range-successor queries at times).  With these data structures and given values $\ell$ and $q$, we can find the largest position of a value less than $\ell$ in $L [0..2 q - 2]$ and the smallest position of a value less than $\ell$ in $L [2 q..2 t]$ in $O (\log^\epsilon t)$ time.

Given the tag statistics $\TS [1..m+1]$ of $P$ with respect to $T$ and $i$ and $j$, we can check that $P [i..j]$ occurs in $T$ at all by verifying that $\TS [i].\len \geq j - i + 1$.  Assuming it does, we can look up the index $q = \TS [i].\run$ of the run in the tag array containing $\Tag [\TS [i].\rank]$ and we can check in constant time whether
\begin{eqnarray*}
\TS [i].\up & \geq & j - i + 1\,,\\
\TS [i].\down & \geq & j - i + 1\,.
\end{eqnarray*}

If $\TS [i].\up  < j - i + 1$ then $L [2 q - 1] < j - i + 1$ (note $L[2q-1]$ is the LCP within run $q$) 
and run $q$ is the first in the tag array to overlap the BWT interval for $P [i..j]$.  Otherwise, we use a range-predecessor query to find the largest position in $L [0..2 q - 2]$ with value less than $j - i + 1$. This tells us the first run in the tag array to overlap the BWT interval for $P [i..j]$: If the range-predecessor query returns $p$, then the index of this first run is $1+\lfloor p/2 \rfloor$; the run is covered completely if $p$ is even and partially if $p$ is odd. 

Symmetrically, if $\TS [i].\down < j - i + 1$ then $L [2 q - 1] < j - i + 1$ and run $q$ is the last one in the tag array to overlap the BWT interval for $P [i..j]$.  Otherwise, we use a range-successor query to find the smallest position  in $L [2 q..2 t]$ of a value less than $j - i + 1$, which tells us the last run in the tag array to overlap the BWT interval for $P [i..j]$. If the range-successor query returns $p$, then the index of this last run is $\lceil p/2 \rceil$, and it is covered completely iff $p$ is even.

Notice we never compute the BWT interval for $P [i..j]$.
\end{proof}

\begin{corollary}
\label{cor:range-successor}
Suppose we are given an SLP with $g$ rules for a text $T [1..n]$ whose $\BWT$ has $r$ runs, and a tag array for $T$ with $t$ runs.  Then, for any constant $\epsilon>0$, we can build an $O (g + r + t)$-space data structure with which we can preprocess any pattern $P [1..m]$ in $O (m \log n)$ time such that later, given $i$ and $j$, we can return the $k$ distinct tags labeling occurrences of $P [i..j]$ in $T$ in $O (\log^\epsilon t + k)$ time.
\end{corollary}

\begin{proof}
We store instances of the data structures from (i)  Corollary~\ref{cor:MONI}, (ii) Lemma~\ref{lem:range-successor}, and (iii) Corollary~\ref{cor:RLBWT}.  Given $P$, we use the data structures (i) to compute the tag statistics of $P$ with respect to $T$ in $O (m \log n)$ time.  Given $i$ and $j$, we use the data structures (ii) to find the indices $s$ and $e$ of the runs in $\Tag$ that are contained in or overlap the BWT range of $P[i..j]$, in time $O (\log^\epsilon t)$. Finally, using the array $A[1..t]$ (iii) we run Muthukrishnan's algorithm on $A[s..e]$ to find the $k$ distinct tags labeling occurrences of $P [i..j]$ in $T$, in $O (k)$ time.
\end{proof}

\section{Optimal-time tag reporting}
\label{sec:optimal}

The time in Corollary~\ref{cor:range-successor} for reporting the $k$ distinct tags labeling occurrences of $P [i..j]$ in $T$---that is, $O (\log^\epsilon t + k)$---is optimal if $k \in \Omega (\log^\epsilon t)$.  We do not know $k$ in advance, however, and if we always want optimal reporting time we cannot afford range-successor queries right away. 

We start with an important property of the ranges we find in $L$ in the proof of Corollary~\ref{cor:range-successor}.

\begin{lemma} \label{lem:hier}
The ranges resulting from range-predecessor/successor queries from positions $q$ in $L$ and thresholds $\ell$ can be equal, disjoint or nested, but cannot overlap.
\end{lemma}
\begin{proof}
Consider one such range $L[u..d]$, which is as large as possible around some $q$ not containing any values less than some $\ell$. Assume there is another range $L[u'..d']$ obtained similarly for position $q'$ and threshold $\ell'$, and that $u < u' \le d < 2t$. It follows that $\ell \le L[u'-1] < \ell'$, therefore, since $L[d+1] < \ell < \ell'$, it must be $d' \le d$, so $L[u'..d']$ is contained in $L[u..d]$. The case $u \le d' < d$ is analogous.
\end{proof}

\begin{figure}[t]
\centering
\includegraphics[width=0.8\textwidth]{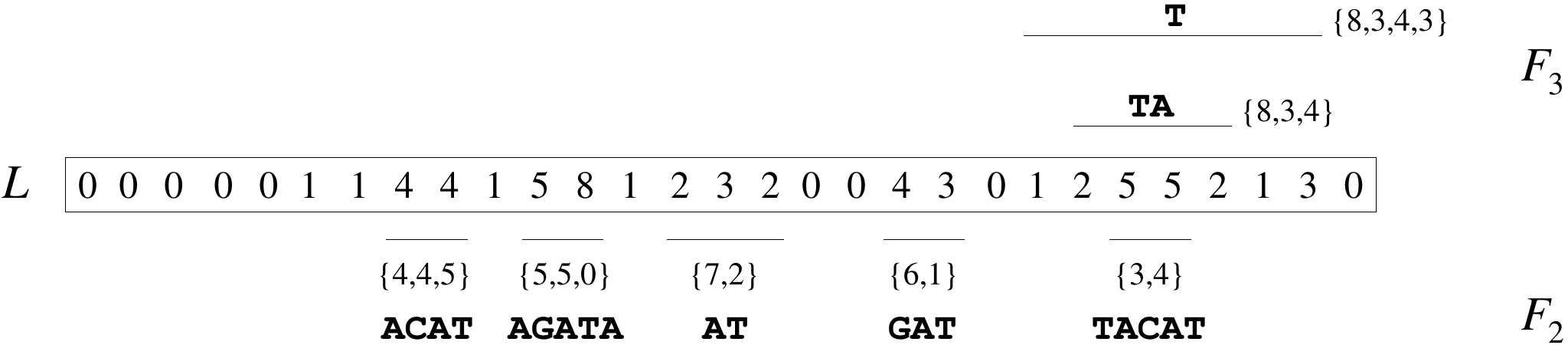}
\caption{The array $L$ of Figure~\ref{fig:rangesuccessor} and the sets of segments forming $F_3$ (above) and $F_2$ (below). The larger range of $F_3$ contains the smaller, and thus they represent the same set of tags.}
\label{fig:F}
\end{figure}

Consider the distinct ranges we can find in $L$ such that the corresponding range in $\Tag$ (including both contained and overlapped runs) contains $k'$ distinct tags, for some $k'$. If two of these ranges in $L$ are nested, then their corresponding ranges in $\Tag$ contain exactly the same $k'$ distinct tags---possibly with different multiplicities, but that does not concern us here. Let $F_{k'}$ be the $O (t)$-bit balanced-parentheses representation \cite{MR01} of these distinct ranges in $L$, where every range is an ancestor of those it contains. With $O(t)$ further bits, we can find in $O(1)$ time the lowest node of $F_{k'}$ that contains any given entry $L[q]$ \cite[Sec.~4.1]{RNO11}. Figure~\ref{fig:F} gives an example.

While querying the data structure from Lemma~\ref{lem:range-successor}, if we somehow guess correctly that our range-successor queries will return a range in $L$ whose corresponding range in $\Tag$ contains exactly $k'$ distinct tags, then we can replace those range-successor queries by the constant-time method described above to find the corresponding node in $F_{k'}$.  

This node may correspond to a range nested strictly inside the one we would obtain from the range-successor queries but, as we noted above, that makes no difference to our final answer.
In fact, the node of $F_{k'}$ we find has the smallest range---corresponding to the largest value of $j - i + 1$---we could obtain from our range-successor queries, while still returning $k'$ distinct tags.  If we store an $O (t)$-bits range-minimum data structure \cite{FH11} over $L$---which we can reuse for all values of $k'$---then we can find that largest value $j - i + 1$ in constant time, as it is the minimum value of $L$ in the range.

Of course, we cannot assume we will guess correctly the number $k'$ of distinct tags we will eventually return.  Instead, we keep an $O(t)$-bits representation $F_{k'}$ for every $k' \leq \lg^\epsilon t$, which takes $O \left( \frac{t \lg^\epsilon t}{\log t} \right) \subset O (t)$ space.  We query $F_1, F_2, F_3, \ldots, F_{\lg^\epsilon t}$ in turn, using constant time for each. If, for some $F_{k'}$, the range-minimum data structure returns a value smaller than $j - i + 1$, then we know that $P[i..j]$ is labeled by $k = k'-1$ distinct tags, so we use the formulas of Section~\ref{sec:using} to convert the range in $L$ given by $F_{k}$ to a range $A[s..e]$, and use Muthukrishnan's algorithm (Corollary~\ref{cor:RLBWT}) to return the distinct tags in $A[s..e]$. Otherwise, after we query $F_{\lg^\epsilon t}$, we know that $k > \lg^\epsilon t$, so we can perform the range-successor queries safely as in Section~\ref{sec:using}. In both cases, we use $O (k)$ total time.


\begin{theorem}
\label{thm:optimal}
Suppose we are given an SLP with $g$ rules for a text $T [1..n]$ whose $\BWT$ has $r$ runs, and a tag array for $T$ with $t$ runs.  Then we can build an $O (g + r + t)$-space data structure that can preprocess any pattern $P [1..m]$ in $O (m \log n)$ time such that later, given $i$ and $j$, it returns the $k$ distinct tags labeling occurrences of $P [i..j]$ in $T$ in optimal $O (k)$ time.
\end{theorem}

\section{Top-$k$ most frequent tags}
\label{sec:topk}

%
%

{Finding matches is only one step in building a consensus sequence that represents our best estimate of a sequenced individual's genome.  Later steps include extending those matches with dynamic programming, chaining, scaffolding and gap-filling.  If we consider too many possible matches, performing these later steps for all of them becomes impractical, so we may want to limit the number of tags our data structure returns.  One obvious approach is to return only the $k$ tags that most frequently label occurrences of $P [i..j]$ in the genomes.} 

Consider now the problem of, given $k$ at query time, retrieving only $k$ tags labeling most occurrences of $P[i..j]$ in $T$, within $O(t)$ space.
This corresponds, in principle, to finding $k$ most frequent tags in $\Tag[s..e]$, where $[s..e]$ is the BWT interval for $P[i..j]$.
Because in Section~\ref{sec:using} we compute this BWT interval only up to the granularity of the runs in $\Tag$, we cannot solve this query precisely because do not know how many times the tag symbol of the first and last run overlapping $\Tag[s..e]$ occurs inside $\Tag[s..e]$. Instead, we will always include those two candidates in the answer and return $k+2$ elements that must contain the top-$k$ tags for $P[i..j]$. 

We then focus in the sequel on finding the top-$k$ most frequent tags in an interval of $\Tag$ formed by whole runs. Section~\ref{sec:using} shows how to obtain the range $A[s'..e']$ of run numbers completely included in the interval $\Tag[s..e]$ for $P[i..j]$. Together with array $A[1..t]$, which stores the tag symbol of each run, we store the run lengths in an array $R[1..t]$. We then say that the weight of $A[i]$ in the $i$th run of $\Tag$ is $R[i]$, and define the weight of a tag $\tau$ in $A[s'..e']$ as 
$\sum_{s' \le i \le e', A[i]=t} R[i]$, that is, the sum of the weights associated with cells containing $\tau$ in $A[s'..e']$. We then recast the problem to finding $k$ tags with maximum weight in $A[s'..e']$.
We assume the interval is nonempty, otherwise the only tags to report are those two in the overlapped runs.

\begin{figure}[t]
\centering
\includegraphics[width=0.7\textwidth]{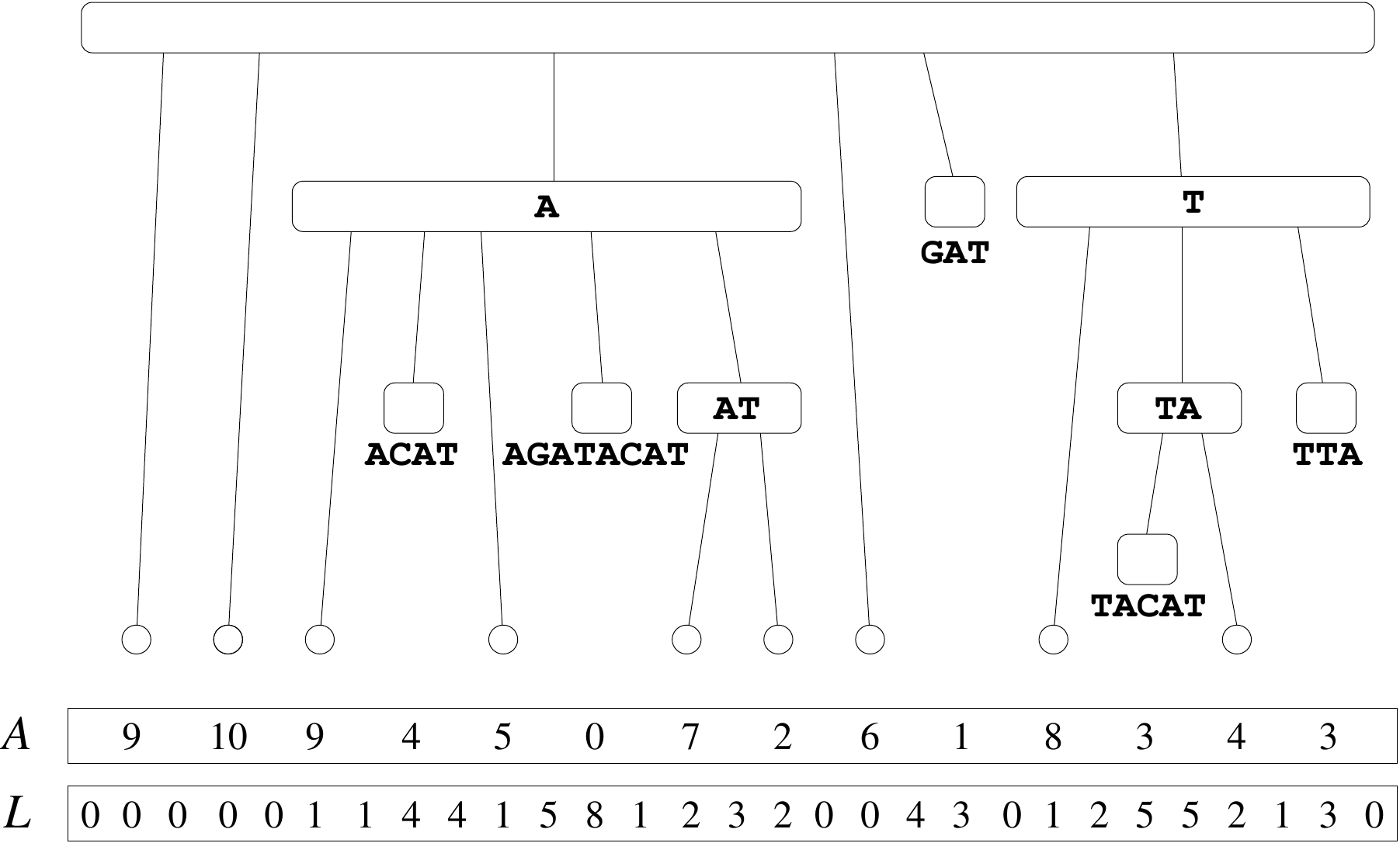}
\caption{The array $L$ of Figure~\ref{fig:rangesuccessor} and the tree $\T$ on top. The small circles are the extra added leaves, and the empty node on top is the special root.}
\label{fig:T}
\end{figure}

While this problem does not have efficient solutions in general (the best solution to finding just the mode in a range takes time $O(\sqrt{n/\log n})$ \cite{CDLMW12}), our case is particular because the set of all the possible queried segments in $A$ forms a containment hierarchy $\T$: they correspond to retaining only the odd positions of array $L$, where the ranges already form a hierarchy by Lemma~\ref{lem:hier}, so the disjointness and containment relations are preserved.
Further, we add a special root as the parent of all the maximal segments, and all the individual odd positions of $L$ as leaves, if they do not already exist, as children of the lowest node that contains their position. Note that $\T$ has $O(t)$ nodes because its nodes correspond to (some) suffix tree nodes. From the positions $s'$ and $e'$, corresponding to leaves in $\T$, we can perform a constant-time lowest common ancestor query \cite{BFCPSS05} to reach the node of $\T$ corresponding to the range $[s'..e']$. Figure~\ref{fig:T} gives an example.

Inspired in ideas from top-$k$ document retrieval \cite{HSTV13,NN17}, we define $\T_\tau$ as the subset of the nodes in $\T$ formed by the special root, the leaves with tag $\tau$, and all the lowest common ancestors of pairs of those leaves. We then define weighted upward {\em pointers} in $\T$ as follows: from every node $v \in \T_\tau$ covering $A[s'..e']$, there is a pointer to its lowest ancestor $u$ in $\T_\tau$; the weight of the pointer is the same weight of $\tau$ in $A[s'..e']$. If there is no such $u$, the pointer leads to the special root node. It is easy to see that there are $O(t)$ pointers, since there is one per node in $\T_\tau$, and the size of $\T_\tau$ is at most twice the number of cells in $A$ with tag $\tau$. 

The key property of this arrangement \cite{HSTV13} is that, for any node $x \in \T$, there is exactly one pointer going from a descendant of $x$ ($x$ included) to an ancestor of $x$ ($x$ not included) per different tag $\tau$ covered by $x$. Therefore, our problem boils down to finding $k$ heaviest such pointers for the lowest common ancestor $x$ of the $s'$th and $e'$th leaves of $\T$. 

We recast this, in turn, as a geometric problem \cite{NN17}. Each pointer going from $v$ to $u$ in $\T$ is regarded as a weighted point $(pre(v),depth(u))$ in a two-dimensional $O(t) \times O(t)$ grid, where the $x$-coordinate $pre(v)$ is the preorder of $v$ in $\T$, and the $y$-coordinate $depth(u)$ is the depth of node $u$ in $\T$. The weight of the point is the same weight of the pointer. Let $min(x)$ and $max(x)$ be the minimum and maximum preorders of nodes in the subtree rooted at $x$, then a pointer from $v$ to $u$ is relevant for $x$ iff $min(x) \le pre(v) \le max(x)$ and $depth(u) < depth(x)$. Our query then boils down to finding $k$ heaviest points in the orthogonal range $[min(x),max(x)] \times [0,depth(x)-1]$. 

This geometric problem can be solved in time $O(\log^{1+\epsilon} t + k(\log\log n + \log^\epsilon t))$ for any constant $\epsilon > 0$ \cite[Lem.~7]{NN17}. The $O(\log\log n)$ term is the cost of the operations on a priority queue handling weights, which are in $[1..n]$. To reduce it, we consider all the distinct weights that can appear in pointers of $\T$. Since there are $O(t)$ pointers overall, we can assign them numbers in $[1..O(t)]$ that respect their original order. The priority queue then operates on a universe of size $O(t)$ and requires time $O(\log\log t)$ per operation, which is absorbed by $O(\log^{\epsilon}t)$.

\begin{theorem}
\label{thm:topk}
Suppose we are given an SLP with $g$ rules for a text $T [1..n]$ whose $\BWT$ has $r$ runs, and a tag array for $T$ with $t$ runs.  Then, for any constant $\epsilon>0$, we can build an $O (g + r + t)$-space data structure that can preprocess any pattern $P [1..m]$ in $O (m \log n)$ time such that later, given $i$ and $j$, and a threshold $k$, it returns at most $k+2$ elements that include $k$ tags labeling the most occurrences of $P [i..j]$ in $T$ in time $O((\log t + k)\log^\epsilon t)$.
\end{theorem}

Finally, to determine which $k$ to choose for this query, it may be useful to know how many distinct tags are there to choose from. This can be easily computed in constant time as follows. Each node of $\T$ covering $A[s'..e']$ stores the number of distinct tags in $A[s'..e']$. It also stores whether (i) the tag $A[s'-1]$ does not appear in $A[s'..e']$, and (ii) the tag $A[e'+1]$ does not appear in $A[s'..e']$. When we find the range of $P[i..j]$ in $L$, from where we obtain the range $A[s'..e']$ of fully contained runs, we also know whether the range partially extends to the $(s'-1)$th and $(e+1)$th runs. The answer is then the number of distinct tags in the node $x \in \T$ that is the lowest common ancestor of its $s'$th and $e'$th leaves, plus 1 if the range intersects the $(s'-1)$th run and (i) holds, plus 1 if the range intersects the $(e'+1)$th run and (ii) holds, minus 1 if we added the two previous points and $A[s'-1] = A[e'+1]$.

\begin{theorem}
\label{thm:count}
Suppose we are given an SLP with $g$ rules for a text $T [1..n]$ whose $\BWT$ has $r$ runs, and a tag array for $T$ with $t$ runs.  Then we can build an $O (g + r + t)$-space data structure that can preprocess any pattern $P [1..m]$ in $O (m \log n)$ time such that later, given $i$ and $j$, it returns the number of distinct tags labeling occurrences of $P [i..j]$ in $T$ in constant time.
\end{theorem}

Note that, if we store the data structures used in this theorem, then it is simpler to obtain Theorem~\ref{thm:optimal}: we first find the number $k$ of distinct tags labeling occurrences of $P[i..j]$.
If $k \le \lg^\epsilon t$, we run the query on the tree $F_k$; otherwise we use the general mechanism of Corollary~\ref{cor:range-successor}.

\section{Constraining occurrence frequencies}
\label{sec:fmems}

As discussed in Section~\ref{sec:topk}, one may be interested in returning only the most frequent tags that match a substring of $P$. In some cases, determining how frequent must a tag be in order to be relevant can be more natural than establishing how many (most frequent) tags we want. 

We now show how to fix
a parameter $f$ at construction time so that we find only the tags that label at least $f$ occurrences of $P[i..j]$ in $T$. This is a natural generalization of the notion of $f$-MEMs \cite[Sec.~11.6.1]{navarro2016compact}. In our solution we build upon the extension of the $r$-index \cite{tatarnikov_et_al2023monik} that allows us to locate the occurrences of any $P[i..j]$ that appears at least $f$ times in $T$ in $O(m \log n)$ time and within $O(r)$ space.
Let us first define some auxiliary data.

\begin{definition}[Triple array]
We use an array $\triple[1..2 t]$ of $(\tag, \ptr, \len)$ triples.  Entries $\triple[2 i - 1]$ and $\triple[2i]$ correspond to the start and end of the $i$-th run in the $\Tag$ array, respectively. $\triple[q].\tag$ stores the tag of the run $\triple[q]$ corresponds to. $\triple[q].\ptr$ stores a pointer (between $0$ and $2 t - 1$) of the preceding entry in $\triple$ with the same tag entry, or $0$ if there is none. Furthermore, if $q$ corresponds to the start (end) of the run $\Tag[u..v]$, then $\triple[q].\len$ is the length of the longest prefix of the suffix $T[\SA[u]..n]$ ($T[\SA[v]..n]$) that occurs in $T$ labeled by $\triple[q].\tag$ at least $f$ times.
\end{definition}

For a pattern $P$ of length $m$, we obtain the tag statistics $\TS[1..m+1]$ and by means of Corollary $\ref{cor:range-successor}$ we find out which runs in $\Tag$ are contained in and overlapped by the BWT interval of $P[i..j]$. We then compute the corresponding interval $\triple[u..v]$, such that we include the entry for the end of the first run in $\Tag$ that overlaps the BWT interval, but exclude the entry for the start of the run (unless the first run is fully contained in the BWT interval); symmetrically, if the BWT interval ends in a $\Tag$ run, then we include the entry for the start of the run but exclude the entry for the end of the run (unless the last run is fully contained). 

We now need to report all the tags that appear at least $f$ times in $\Tag[u..v]$. Naturally, for $\tau \in [u, v]$ we report $\triple[\tau].\tag$ whenever $\triple[\tau].\len \geq j-i+1$. To avoid reporting multiple copies of the same tag, the distinct tags to be reported are uniquely mapped to entries in $\triple[u..v]$ in a way similar to Muthukrishnan's document listing \cite{muthukrishnan2002efficient}, which is summarized in the following lemma.

\begin{lemma}
\label{lem:3Dpoint}
Let $P[i..j]$ be a substring of $P$ with the corresponding interval $\triple[u..v]$. If $P[i..j]$ occurs labeled by a tag $\tau$ at least $f$ times in $T$, then there is exactly one entry $\triple[k]$ such that $k \in [u, v]$, $\triple[k].\tag = \tau$, $\triple[k].\ptr < u$ and $\triple[k].\len \geq j-i+1$. 
\end{lemma}

\begin{proof}
Consider the smallest index $k \in [u,v]$ such that $\triple[k].\tag = \tau$. The existence of such $k$ is follows from the fact that $P[i..j]$ occurs in $T$ labeled by $\tau$. Since $k$ is the first entry in $\triple[u..v]$ that with the tag $\tau$, it follows that $\triple[k].\ptr < u$. Lastly, $\triple[k].\len \geq i-j+1$ since at least $f$ occurrences of $P[i..j]$ in $T$ are labeled by $\tau$.
\end{proof}

To find all the positions in $\triple$ characterized by Lemma \ref{lem:3Dpoint} and report associated tags, we will build a 3-D range query data structure of points $(k, \triple[k].\ptr, \triple[k].\len)$ with $\triple[k].\tag$ as satellite data.

We proceed by considering whether the BWT interval for $P[i..j]$ spans at least two runs in $\Tag$ or not. If it does, it is sufficient to perform a $4$-sided range query on $\left[u, v\right] \times \left[0,u-1\right] \times \left[i-j+1, n\right]$ and report all the distinct tags of the associated points. If the BWT interval of $P[i..j]$ is contained within a single run in $\Tag$, we use the solution by Tatarnikov et al.~\cite{tatarnikov_et_al2023monik} to obtain the $f$-matching statistics, we test whether a sufficiently long prefix of $P[i..m]$ occurs more than $f$ times in $T$ and, depending on the answer, report the single tag or nothing. This method accompanied by a novel data structure by Nekrich \cite{nekrich2021new} to answer 3-D orthogonal range queries yields the following theorem.

\begin{theorem}
Given a text $T[1..n]$ and a parameter $f$, we can construct for any constant $\epsilon>0$ a data structure of size $O (r + g + t)$ with we can preprocess a pattern $P$ of length $m$ in $O (m \log n)$ time, such that later, given $i$ and $j$, it returns the $k$ distinct tags that each label at least $f$ occurrences of $P[i..j]$ in $T$ in $O((k+1)\log^\epsilon t)$ time. 
\end{theorem}

\section{Discussion and future work} \label{sec:concl}

This paper lays out the theoretical basis for Wheeler maps, and we are now investigating them experimentally.  As well as full implementations of the data structures described here, we also still need efficient algorithms for extracting tag arrays from pangenome graphs for genomic datasets, and good compression schemes for those tag arrays.  A tag could contain a lot of information, so representing it explicitly for every run of that tag in the tag array might be very wasteful.  It is likely more space efficient to store each distinct tag only once, separated from the tag array by one or more levels of indirection.

Once we can build and store Wheeler maps well in practice, we intend to integrate them into current pangenomics pipelines.  Of course, we will also look for other applications of Wheeler maps.  It may even be that, in some circumstances, mixing Wheeler graphs with Wheeler maps --- to allow some kinds of recombinations while excluding others --- is useful.  Finally, we will look for cases in which we can prove analytical bounds on $t$, trying not to overconstrain the structure of the Wheeler map.

\bibliography{references.bib}

\end{document}